\documentstyle[aps,eqsecnum]{revtex}

\begin{document}

\title{\bf Evidence for the adiabatic invariance of the black hole
  horizon area}

\author{Avraham E. Mayo
\thanks{Electronic adsress :Mayo@shum.cc.huji.ac.il}}
\address{\it The Racah Institute of Physics, Hebrew University
of Jerusalem,\\ Givat Ram, Jerusalem 91904, Israel}
\date{\today}
\maketitle

\begin{abstract}
Some examples in support of the conjecture that the horizon area
of a near equilibrium black hole is an adiabatic invariant are
described. These clarify somewhat the conditions under which the
conjecture would be true.
\end{abstract}


\section{Introduction and Summary}
Consider the event horizon area of a black hole. Hawking's area
theorem \cite{AreaTheorem} indicates that whenever classical fields
obeying the weak energy condition are involved, this physical
entity must always grow under external disturbances. Be that as it
may, it is possible to exhibit an assortment of cases in which the
slow application and relaxation of an external disturbance does not
necessarily mandate an area increase. This is of course in
agreement with the identification of the horizon area and the
entropy associated with the black hole \cite{AreaEntropy}. After
all, in classical thermodynamics entropy is invariant under slow,
reversible changes of a system in thermodynamic equilibrium. This
implies a similarity of horizon area to an adiabatic invariant in
mechanics. What is an adiabatic invariant?

Consider a Hamiltonian system characterized by a Hamiltonian
function $H(p,q;\lambda)$ depending on a parameter $\lambda$.
Suppose that the parameter changes slowly with time. It turns out
that in the limit as the rate of change of the parameter approaches
$0$, there is a remarkable asymptotic phenomenon: two quantities,
generally independent, become functions of each other. For
example, consider the motion of a perfectly elastic rigid ball
between perfectly elastic walls whose separation $l$ slowly varies.
In this case the product $vl$ of the velocity of the ball $v$ and
the distance between the walls turns out to be an adiabatic
invariant, e.g. if we increase very slowly the separation and then
very slowly decrease it to the original value (slow in comparison
with the characteristic oscillation), then at the end of this
process the velocity of the ball will be the same as it was at the
start. Additionally, it turns out that the ratio of the energy $H$
of the ball to the frequency of the oscillation between the walls
$\omega$ changes very little under a slow change of the parameter,
although the energy and frequency themselves may change a lot.
Quantities such as this ratio are called {\it adiabatic invariants}.
A more precise definition of adiabatic invariant is \cite{Arnold}

{\bf Definition.} The quantity $I(p,q;\lambda)$ is an {\it adiabatic
  invariant} of the system
\begin{equation}
\stackrel{\cdot}{p}=-{\partial H\over \partial q},\quad\quad
\stackrel{\cdot}{q}={\partial H\over \partial p},\quad\quad
H=H(p,q;\epsilon t).
\end{equation}
if for every $\kappa>0$ there is an $\epsilon_0>0$ such that if
$\epsilon<\epsilon_0$ and $0<t<1/\epsilon$, then
\begin{equation}
\left|I(p(t),q(t);\epsilon t)-I(p(0),q(0);0)\right|<\kappa.
\end{equation}
Note that this definition allows for the preservation of the
adiabatic invariant with a power-law accuracy (with respect to the
small parameter describing the speed of frequency change), in
comparison with the exponential accuracy considered to be generic
\cite{LL}.

In quantum terms, one may comprehend the adiabatic invariance of
$H/\omega$ by the following arguments. Since it is assumed that
during the adiabatic change of the system the perturbations imposed
have frequencies very small compared to $\omega$, transitions
between states of successive quantum numbers are strongly
suppressed. Now, for example, for a harmonic oscillator in a
stationary state labeled by quantum number $n$, $H/\omega = (n+{1/
2})\hbar$. Therefore, the ratio $H/\omega$ is preserved.

A general proof that the horizon area is an ``adiabatic invariant''
will enable the utilization of Ehrenfest's principle~\cite{Eh}: any
classical adiabatic invariant corresponds to a quantum entity with
discrete spectrum. Ehrenfest  showed that for a quasiperiodic system,
all Jacobi action integrals of the form $I\,=\,\oint p\, dq$ are
adiabatic invariants. This principle is the backbone of the
Bohr-Sommerfeld theory- ``old quantum mechanics'', where all Jacobi
actions, invariant under adiabatic changes, are quantized in
integers.

Even so, application of this principle to the problem at hand or
to any other problem does not depend on whether the adiabatic
invariant can be expressed as a Jacobi action integral or not. In
fact, since the Hamilton-Jacobi formalism for black holes is not
very well developed, it is not even clear that the horizon area of a
black hole can be so expressed. First attempts in this direction were
made in the framework of the membrane paradigm \cite{membrane}. In
this framework the problem of slow time dependent perturbations of a
black hole is the basis for the membrane description of black hole
evolution. This describes the horizon as a viscous conductive two
dimensional surface. Recently Parikh and Wilczek \cite{Parikh} have
calculated an action for black hole membranes using the Euclidean
formulation of Teitelboim \cite{Teitelboim}. Hence, as matters stand
we find it useful to seek insights into the problem using a different
approach which is based on the Newman-Penrose formalism as presented in
the next section.

Accordingly, application of Ehrenfest's principle will then lead
to the spectrum for the horizon area operator. On that account, the
object of this work is to present a collection of examples in which
the horizon area of the black hole shows signs of being the analogue
of a mechanical adiabatic invariant (not necessarily in the form of a
Jacobi action integral). Each example is decomposed into a static and
dynamical problem. As can be easily understood, in the static case
there cannot be any energy flux through the horizon, and thus no area
increase. However, the static case
serves as a limiting case for the dynamical problem as we take the
adiabatic parameter to zero (it is not clear beforehand that this
limit actually exists). These examples suggest the existence of a
theorem which would state that, classically, under suitably adiabatic
changes of a black hole in equilibrium, the area of its event horizon does not
change. This would provide a formal motivation for quantizing the
black hole in the spirit of the ``old quantum theory''. This approach
has lately received increased attention \cite{BekMkhanov}.

The paper is organized as follows. In Sec.~\ref{idea} we present
the basic theory required to answer the question regarding the
circumstances under which the black hole horizon area remains
unchanged under the influence of an exterior perturbation. In
Sec.~\ref{RN1} we discuss the generalization of the result obtained
by Bekenstein \cite{BekBrazil3} regarding the influence of scalar
charges in the Schwarzschild black hole exterior on the horizon's
area to the case in which the black hole is electrically charged.
The extreme case of the previous example is discussed in
Sec.~\ref{Ern}. As a further example we consider in Sec.~\ref{d-S}
the possibility that the charges are distributed inside a de Sitter
universe with a Schwarzschild black hole embedded inside it. The
influence of the perturbation on the cosmological horizon will be
also discussed. In Sec.~\ref{kerr} we study in brief the
possibility that the hole itself is rotating. Finally, in
Sec.~\ref{electro} we examine the influence of an electromagnetic
radiation on the horizon area of a Schwarzschild black hole.

\section{The Basic Idea}
\label{idea}
Here we shall lay out the basic theory needed to answer the
question regarding the circumstances under which the black hole
horizon area remains unchanged under the influence of an exterior
perturbation.

Consider a small patch of event horizon area $\delta A$; it is
formed by null geodesic generators whose tangents vectors are
$l^\alpha=dx^\alpha/d\lambda$, where $\lambda$ is an affine
parameter along the generators \cite{HawkHartle}. The complex
conjugate null vectors $m^\alpha$ and $\overline{m}^\alpha$
together with $l^\alpha$ and a fourth real vector $n^\alpha$
orthogonal to them which satisfies $l^\alpha n_\alpha=1$,
constitute the Newman--Penrose tetrad which lies at the horizon.
$m^\alpha$ and $\overline{m}^\alpha$ are orthogonal to $l^\alpha$
and satisfy $m^\alpha\overline{m}_\alpha=-1$.

By definition of the convergence of the generators $\rho$, $\delta A$
changes at a rate
\begin{equation}
{d\over d\lambda}\delta A = -2\rho\delta A.
\label{changeA}
\end{equation}

Now the convergence, defined by $\rho\equiv l_\alpha ; _\beta
m^\alpha\overline{m}^\beta$, changes at a rate given by the optical
analogue of the Raychaudhuri equation \cite{NP,Pirani}
\begin{equation}
{d \rho\over d\lambda} = \rho^2 + \sigma\overline{\sigma} + 4\pi
T_{\alpha\beta}\,l^\alpha l^\beta,
\label{changerho}
\end{equation}
where $\sigma\equiv l_\alpha ; _\beta m^\alpha m^\beta$ is the shear
of the generators and $T_{\alpha\beta}$ is the energy momentum tensor
of the matter in the exterior of the black hole. The shear, which
measures deformation of the horizon,  evolves according to
\begin{equation}
{d\sigma\over d\lambda} = 2\rho\sigma +C_{\alpha\beta\gamma\delta}\,
l^\alpha m^\beta l^\gamma \overline{m}^\delta,
\label{changesigma}
\end{equation}
where $C_{\alpha\beta\gamma\delta}$ is the Weyl conformal tensor
\cite{MTW}. Utilizing the orthogonality relations of the
tetrad base and Einstien's equations one may calculate the Weyl
conformal tensor on the horizon
\begin{equation}
  \label{weyl_on_horizon}
   C_{\alpha\beta\gamma\delta}\,l^\alpha m^\beta l^\gamma
   \overline{m}^\delta=R_{\alpha\beta\gamma\delta}\,l^\alpha m^\beta
   l^\gamma \overline{m}^\delta+
4\pi  T_{\alpha\beta}\,l^\alpha l^\beta.
\label{C=R+T}
\end{equation}

It will be noticed that to keep the horizon area constant requires
$\rho=0$ which by Eqs.~(\ref{changerho}) (\ref{changesigma}) implies
that both $C_{\alpha\beta\gamma\delta}\, l^\alpha m^\beta l^\gamma
 \overline{m}^\delta $ and $T_{\alpha\beta}\,l^\alpha l^\beta $ vanish at
the horizon. Thus, if initially the convergence and shear were
equal to zero, then they would remain so. Vanishing of
$C_{\alpha\beta\gamma\delta}\, l^\alpha m^\beta l^\gamma \overline{m}^\delta
$ requires that the geometry be quasistationary to prevent
gravitational waves, which are quantified by
$C_{\alpha\beta\gamma\delta}$, from impinging on the horizon. Thus
with a quasistationary geometry, preservation of the horizon area
requires
\begin{equation}
T_{\alpha\beta}\,l^\alpha l^\beta  = 0 \quad\quad {\rm on\ the\ horizon}.
\label{zerocondition}
\end{equation}

Note that in the dynamical case one should expect a total area increase
which can be calculated by first integrating Eqs.~(\ref{changerho}) 
(\ref{changesigma}) and then substituting the result in
Eq.~(\ref{changeA}). However, since we are mainly interested in
the adiabatic character of the horizon area it is sufficient to prove
that the rate of change of the horizon area, given
by $(d\,\delta A/d\lambda)/\delta A$ is smaller than the rate of
deformation, given by $\sigma$, (note that the comparison is between
two quantities with the same dimensions). Now, according to
Eq.~(\ref{changeA}) $(d\,\delta A/d\lambda)/\delta A$ is
proportional to $\rho$. Hence, it is sufficient to prove that the
differential change in $\rho$ is smaller than the differential change
in $\sigma$. Evidently the entire evolution of the various quantities in
the problem is governed by the magnitude of the Riemann tensor and the
energy-momentum tensor on the horizon. Therefore, according to
Eqs.~(\ref{changerho}) (\ref{changesigma}) all that is left to verify
in order to prove our claim is that the term
$T_{\alpha\beta}\,l^\alpha l^\beta$ is of higher order in the
adiabatic parameter than the term
$R_{\alpha\beta\gamma\delta}\,l^\alpha  m^\beta l^\gamma
\overline{m}^\delta$.

 A quick way to see that it is indeed a true claim goes as
follows. The distribution of static perturbators around a static black
hole should not cause an area increase (this is but a simple
consequence of staticity). This can be realized by noting that the
zeroth order term in the adiabatic parameter of
$R_{\alpha\beta\gamma\delta}\,l^\alpha  m^\beta l^\gamma
\overline{m}^\delta$ must vanish. However, in general the first order
term would not vanish. For example, such terms appear when a
contraction is made between the first order correction to the Riemann
tensor and the zeroth order corrections to the tetrad base or {\it
  vice versa}. On the other hand, the contraction of the
energy-momentum tensor with the generators of the horizon must be an
{\it even} function of the adiabatic parameter. This claim is based on
the demand that this term should be a semi-positive definite quantity
as indicated by the weak energy condition, which is assumed to be
obeyed by most of the classical forms of matter found in nature. The
lowest order available is of course the second one. 

Contrary to the common belief which necessitates an increase of horizon
area when changes in the black hole take effect, we shall here
examine a variety of processes for which the conditions that keep the
 horizon area unchanged occur naturally. The rule that seems to
emerge is that quasistationary changes of the black hole induced by
an external influence will leave the horizon area preserved. This
means an ``adiabatic theorem'' for black holes must exist.

\section{Reissner-Nordstr\"om Black Hole Disturbed by Scalar Charges}
\label{RN1}
In this section we extend the result obtained by Bekenstein
\cite{BekBrazil3} regarding the influence of scalar charges in the
Schwarzschild black hole exterior on the horizon area to the
Reissner-Nordstr\"om case. As mentioned already in the
introduction, this example, as for the following examples, is
decomposed into a static and dynamical problem. As can be easily
understood, in the static case there cannot be any energy flux
through the horizon, and thus no area increase. However the static
case serves as a limiting case for the dynamical problem as we take
the adiabatic parameter to zero.

\subsection{The Static Problem}
Consider a Reissner-Nordstr\"om black hole with the metric
\begin{equation}
ds^2 = - (1-2M/r+Q^2/r^2) dt^2 + (1-2M/r+Q^2/r^2)^{-1} dr^2
 + r^2 (d\theta^2 + \sin\theta^2 d\varphi^2).
\label{RN}
\end{equation}
The horizon is located at $r_{\cal H}=M+\sqrt{M^2-Q^2}$. If the
scalar's sources are weak, one may regard the scalar field, $\Phi$,
as a quantity of first order, and conduct a perturbation analysis.
The scalar's energy--momentum tensor,
\begin{equation}
T_\alpha^\beta=\nabla_\alpha\Phi\nabla^\beta\Phi
-{1\over2}\delta_\alpha^\beta\nabla_\gamma\Phi\nabla^\gamma\Phi
\label{EMTS}
\end{equation}
will be of second order of smallness. Thus to first order the
metric (\ref{RN}) is preserved.

To begin with we shall consider first the case in which the scalar
charges are static. Then, the field equation outside the scalar's
sources can be written in the form
\begin{equation}
{\partial\over\partial r} \left( (r^2-2Mr+Q^2 ) {\partial\Phi\over
\partial r}\right) - \hat L^2\,\Phi = 0,
\label{scalarequation}
\end{equation}
where  $\hat L^2$ is the usual squared angular momentum
operator. Following an almost identical derivation as in
\cite{BekBrazil3} for the Schwarzschild black hole case,
one finds the following {\it physical} solution
\begin{equation}
\Phi = \Re \sum_{\ell=0}^\infty\, \sum_{m=-\ell}^\ell
\,C_{\ell m}\, P_\ell\left({r-M\over\sqrt{M^2-Q^2}}\right)\,
Y_{\ell m}(\theta,\varphi)
\label{newPhi}
\end{equation}
where the $Y_{\ell m}$ are the familiar spherical harmonic and
$P_\ell(x)$ are the well-known Legendre polynomials
\cite{MathewsWalker}. The (complex) coefficients $C_{\ell m}$
permit us to match the solution to every distribution of sources by
the usual methods.

An explanation of the term ``{\it physical} solution'' is in order.
For the solution to be physically acceptable, $\Phi$ must not
induce any divergences in the invariants of the curvature via
Einstein's equations. This means that the invariant
$\nabla_\alpha\Phi\nabla^\alpha\Phi$ must be bounded everywhere
(any other invariant is proportional to some power of this one).
Hence, if the other independent solutions of
Eq.~(\ref{scalarequation}), furnished by the Legendre associated
functions $Q_\ell(x)$ (which are singular for $x=0$) were kept,
this condition would be violated.

Keeping in mind that we are interested in the behavior of the
expression $T_{\alpha \beta} l^\alpha l^\beta$ near the black
hole's horizon, we turn now to determine the components of the
vector $l^\alpha$ in the new geometry. Recall that $l^\alpha$ is
the tangent to the null generator of the horizon. Furthermore,
$l^\alpha$ is future directed as well as outgoing. Actually, there
is no need for determination of the components of $l^\alpha$. Any
3D--hypersurface of the form $\{\forall t, r={\rm const}\}$ has a
tangent $\tau^\alpha = \delta_t{}^\alpha$ with norm
$-(1-2M/r+Q^2/r^2)$ as well as the normal $\eta_\alpha =
\partial_\alpha (r-{\rm const}) =\delta_\alpha{}^r$ with norm
$(1-2M/r+Q^2/r^2)$. The linear combination of these two vectors,
$N^\alpha\equiv \tau^\alpha+(1-2M/r+Q^2/r^2) \eta^\alpha$ is
obviously null, and as $r\rightarrow r_{\cal H}$, both its
covariant and contravariant forms remain well defined. Hence, it
must be proportional to $l^\alpha$. Note that just like $l^\alpha$,
$N^\alpha$ is future directed as well as outgoing. Applying the new
found $N^\alpha$ and substituting the energy-momentum tensor for
the scalar field, Eq.~(\ref{EMTS}) one is able to write
\begin{equation}
\lim_{r\rightarrow r_{\cal H}} T_{\alpha\beta} N^\alpha
N^\beta=\lim_{r\rightarrow r_{\cal H}}
(N^\alpha\nabla_\alpha \Phi)^2=\lim_{r\rightarrow r_{\cal H}}
\left((1-2M/r+Q^2/r^2)\partial\Phi/\partial r\right)^2= 0 .
\label{RNTNN}
\end{equation}
Equation (\ref{RNTNN}) implies that the horizon area is left unaffected
by the presence of the static perturbations in the outer region of
a Reissner-Nordstr\"om black hole. This is essential if the area
is an adiabatic invariant under dynamical perturbations of the
black hole as is shown in the next subsection.

\subsection{The Time Dependent Problem}
The time dependent scalar equation in the background
Reissner-Nordstr\"om spacetime is
\begin{equation}
-{r^4\over (r^2-2Mr+Q^2)}{\partial^2\Phi \over \partial t^2}+
{\partial\over \partial r} \left( (r^2-2Mr+Q^2)
{\partial\Phi\over\partial r} \right) - \hat L^2\,\Phi = 0.
\label{tdependence}
\end{equation}
In analogy with Eq.~(\ref{newPhi})  we now look for a solution of
the form
\begin{equation}
\Phi = \Re \int_{0}^{\infty} d\omega\sum_{\ell=0}^\infty
\sum_{m=-\ell}^\ell C_{\ell m}(\omega)\, f_{\ell}(\omega,r)\,
Y_{\ell m}(\theta,\varphi) e^{-\imath\omega t}.
\label{Phitd}
\end{equation}
In terms of Wheeler's generalized ``tortoise'' coordinate
\cite{MTW}, defined by  $dr^*/dr = (1-2M/r+Q^2/r^2)^{-1}$ for which
the horizon resides at $r^* = -\infty $,  the equation satisfied by
the new radial function  $H_{\ell\omega}(r^*)\equiv
rf_\ell(\omega,r)$ is a Schr\"odinger type equation
\begin{eqnarray}
&& \left(-{d^2\over dr^{*2}} + V_\ell(r^*)\right)H_{\ell\omega}=
\omega^2 H_{\ell\omega}
\nonumber \\
&& V_\ell(r^*)=\left(1-{2M\over r}+{Q^2\over r^2}\right)
\left({2M\over r^3} -{2Q^2\over r^4}+ {\ell(\ell+1)\over r^2}\right).
\label{Feq}
\end{eqnarray}
Here $V_\ell(r^*)$, with $r$ considered a function of $r^*$, is a
curvature induced potential barrier \cite{Bicak}. The exact
position of the peak depends on $M$, $Q$ and $\ell$ in general.
When $Q=0$, the peak occurs between $r=3M/8$ $(\ell=0)$, and $r=3M$
$(\ell\rightarrow\infty)$; when $|Q|=M$ the maximum is at $r=2M$
for all $\ell$. The height of the potential is inversely
proportional to $M^2$: for $Q=0$, the maximum is
$V_{0,max}=27/(1024M^2)$, $V_{\ell ,max}=\ell(\ell+1)/(27M^2)$ as
$\ell\rightarrow\infty$; for $|Q|=M$, $V_{\ell
,max}=(1+2\ell(\ell+1))/(32M^2)$. Note that as we approach the
horizon the potential decays exponentially with a vanishing value on
the horizon.

The analogy between Eq.~(\ref{Feq}) and the Schr\"odinger eigenvalue
equation calls for the following analysis of the effects of
distant scalar sources on the black hole horizon  \cite{MTW}. Waves with
``energy'' $\omega^2$ on their way in from a distant source run into
 a positive potential barrier. Therefore, waves with any $\ell$ and
 $\omega^2$ smaller then the maximal hight of the barrier, coming from
 sources at $r\gg 3M$, have to tunnel through the potential barrier to
 get near the horizon. In due course, the wave amplitudes that
 penetrate to the horizon are small fractions of the initial
 amplitudes, most of the waves being reflected back. This means that
 adiabatic perturbations by distant sources perturb the horizon very
 weakly. Thus one would not expects significant growth of the horizon
area from scalar perturbations originating in distant sources. 

What if the scalar's sources are moved into the region $r_{\cal H} < r
< 3M$ inside the barrier? They will now be able to perturb the
horizon; do they change its area? To answer these questions, we look
for the solutions of Eq.~(\ref{Feq}) in the region near the horizon
where the potential is small compared to $\omega^2$.

But before undertaking this task one should note that as $\omega$
tends to zero the sources must also be moved closer and closer to
the horizon since the inside of the potential barrier moves in
towards the horizon. This suggests that there are in fact two time
scales in the problem; one due to the time scale of the sources and
a second one due to the rate of change of their position. However,
since the potential decays exponentially as we come closer and closer
to the horizon this means that we can lower $\omega$ without changing
the location of the sources very much.

Consequently  we turn to find the solutions to Eq.~(\ref{Feq}). According to
the theory of linear second order differential equations they are of the form
\begin{equation}
H_{\ell\omega}(r^*) = \exp(\pm \imath\omega r^*)\times[1 +
O(1-2M/r+Q^2/r^2)].
\label{solution}
\end{equation}
The Matzner boundary condition \cite{Matzner} that the physical
solution be an ingoing wave as appropriate to the absorbing character
of the horizon selects the sign in the exponent as negative. Hence
the typical term in $\Phi$ is
\begin{equation}
{1 +  O(1-2M/r+Q^2/r^2)\over r} P_\ell(\cos\theta)\,\cos\psi;\qquad\qquad
\psi\equiv
\omega(r^*+t)-m\varphi.
\label{term}
\end{equation}

As can be easily verified, invariants of the curvature associated with
this solution [all proportional to
$\left(\nabla_\alpha\Phi\nabla^\alpha\Phi\right)^k$] are bounded. For
example, consider a $\Phi$ composed of a single mode as in
Eq.~(\ref{term}). An explicit calculation on the Reissner-Nordstr\"om
background using $dr^*/dr = (1-2M/r+Q^2/r^2)^{-1}$ gives
\begin{equation}
\nabla_\alpha\Phi\nabla^\alpha\Phi\propto
{m^2 P_\ell{}^2\sin^2\psi\over r^4\sin^2 \theta}
 +\left({dP_\ell\over d\theta}\right)^2{\cos^2\psi\over r^4}
+{\omega \sin(2\psi)\over r^3}P_\ell^2 + \cdots\, ,
\label{miracle}
\end{equation}
where ``$\ \cdots\ $'' here and henceforth denote terms that vanish
as $r$ goes to $r_{\cal H}$. This expression is obviously bounded
at the horizon. By induction it is possible to show that a $\Phi$
which is the sum of several modes is bounded on the horizon.

Finally we turn our attention to the question whether the horizon
area is left unchanged by the perturbation. As with the static case, the
expression which governs the extent by which the horizon area
increases is proportional to
\begin{equation}
T_{\alpha \beta} N^\alpha
N^\beta=\left(\partial\Phi/\partial t+(1-2M/r+Q^2/r^2)
\partial\Phi/\partial r\right)^2
\label{TNNrn}
\end{equation}
If one now substitutes a $\Phi$ made up of a single mode like in
Eq.~(\ref{term}), one finds out that
\begin{equation}
T_{\alpha\beta} l^\alpha l^\beta \propto
{\omega^2 P_\ell^2\sin^2\psi\over r^2} + \cdots\, .
\label{term2}
\end{equation}

In conclusion, on the one hand the amount by which the {\it
shape\/} of the horizon is perturbed when the scalar field sources
are moved inside the potential barrier does not vanish as the
perturbation changes slowly. This quantity is measured by the
complex shear as defined in the previous section. According to
Eqs.~(\ref{changerho})(\ref{changesigma}), the evolution of the
shear is governed by the value of the Weyl tensor on the horizon.
This can be expressed by the Riemann tensor and the energy momentum
tensor on the horizon [see Eq.~(\ref{C=R+T})]. Now, the metric
perturbation, via Einstein's equations, in general must be linear
in the magnitude of the invariant $\nabla_\alpha\Phi\nabla^\alpha\Phi$
(after all Einstein's equations have $T_{\alpha\beta}$ as source, not
$T_\alpha{}^\gamma T_\gamma{}^\beta$). By Eq.~(\ref{miracle}) this
perturbation is of order $O(\omega^0)$ generically, and of $O(\omega)$
in the monopole case. Clearly, the zeroth order of perturbation $O(\omega^0)$,
corresponds to the static case which was already dealt with
previously. Therefore, the {\it dynamic deformation} (to be compared with
{\it static deformation}) is of order $O(\omega)$. On the other hand we have
shown that the rate of change of the horizon area is of order
$O(\omega^2)$, so that as $\omega\rightarrow 0$ the change in area
becomes entirely negligible on the scale of the perturbation.

\section{Extreme Reissner-Nordstr\"om Black Hole Disturbed by Scalar
  Charges}\label{Ern}
It is a well-known fact that the area of an extreme
Reissner-Nordstr\"om black hole is not invariant under the
absorption of a  point particle \cite{BekMkhanov,BekBrazil3}. But
as we shall prove now the horizon area of an {\it extreme\/}
Reissner-Nordstr\"om black hole is indeed an adiabatic invariant
under scalar field-black hole interaction.

\subsection{The Static Problem}
The scalar field equation in the extreme  Reissner-Nordstr\"om
background for a time independent situation is
\begin{equation}
{\partial\over \partial r} \left( (r-M)^2 {\partial\Phi\over\partial r}
\right) - \hat L^2\,\Phi = 0
\end{equation}
with the general solution
\begin{equation}
\Phi = \Re \sum_{\ell=0}^\infty\, \sum_{m=-\ell}^\ell
\,\left({A_{\ell m}\over(r-M)^{\ell +1}}+B_{\ell m}(r-M)^\ell\right)\,
 Y_{\ell m}(\theta,\varphi).
\end{equation}
The complex coefficients $A_{\ell m}$ and $B_{\ell m}$ enable us to
match the solution to any distributions of sources by the usual
methods. Now, since we are in search of a solution which is regular on
the horizon [recall that invariants such as $(\nabla_\alpha\Phi
\nabla^\alpha\Phi)^k$ must be well behaved there], we must make all the
coefficients $A_{\ell m}$ vanish. Thus in the inner region of the
black hole exterior (inwards from the sources) we have, to first
order in perturbations theory, the exact solution
\begin{equation}
\Phi = \Re \sum_{\ell=0}^\infty\, \sum_{m=-\ell}^\ell
\,B_{\ell m}(r-M)^\ell\, Y_{\ell m}(\theta,\varphi).
\end{equation}

As with the non-extreme case we shall employ the vector $N^\alpha$.
In the new geometry, $N^\alpha=\tau^\alpha+(1-M/r)^2 \eta^\alpha$.
The change in the horizon area as measured by
\begin{equation}
\lim_{r\rightarrow M} T_{\alpha \beta} N^\alpha N^\beta=
\lim_{r\rightarrow M}(N^\alpha\nabla_\alpha\Phi)^2=\lim_{r\rightarrow
M}\left((1-M/r)^2 \partial\Phi/\partial r\right)^2
\end{equation}
obviously vanishes. Hence, the horizon area is left unaffected by
the perturbation.

\subsection{The Time Dependent Problem}
When the sources are moved slowly from infinity into the immediate
surroundings of the black hole horizon, we must retain the time
derivatives in the scalar field equation
\begin{equation}
-{r^4\over (r-M)^2}{\partial^2\Phi \over \partial t^2}+
{\partial\over \partial r} \left( (r-M)^2 {\partial\Phi\over\partial r}
\right) - \hat L^2\,\Phi = 0.
\end{equation}
In analogy with the derivations in the previous sections, one should
look for solution of the form
\begin{equation}
\Phi = \Re \int_{0}^{\infty} d\omega\sum_{\ell=0}^\infty
\sum_{m=-\ell}^\ell C_{\ell m}(\omega)\, f_{\ell}(\omega,r)\,
Y_{\ell m}(\theta,\varphi) e^{-\imath\omega t}.
\end{equation}
As usual, it is possible to transform the last equation  with the
aid of the transformations $r^*=r+2M\ln(r/M-1)-{M/(r/M-1)}$ and
$H_{\ell \omega}(r^*)=rf_\ell(\omega,r)$ into
\begin{eqnarray}
&& \left(-{d^2\over dr^{*2}} + V_\ell(r^*)\right) H_{\ell\omega}=
\omega^2 H_{\ell\omega},
\nonumber \\
&& V_\ell(r^*)\equiv\left(1-{M\over r}\right)^2\left({2M\over r^3}
-{2M^2\over r^4}+ {\ell(\ell+1)\over r^2}\right).
\end{eqnarray}
As in the non-extremal case, when the sources are moved through the
potential barrier into the region where $\omega ^2$ is much greater
than the potential, one can look for a solution of the form
\begin{equation}
H_{\ell\omega}(r^*) = \exp(\pm \imath\omega r^*)
\times[1 +  O(1-M/r)^2].
\end{equation}
Again, the Matzner boundary conditions select the sign in the
exponent as negative.  Hence, a typical term in $\Phi$ is as in
Eq.~(\ref{term}) (with M equal to Q). As with the non-extreme case it
can be easily verified that the invariant
$\nabla_\alpha\Phi\nabla^\alpha\Phi$ is bounded on the horizon, and is
generally of order $O(\omega^0)$. Finally the term
$T_{\alpha\beta}N^\alpha N^\beta$ reads
\begin{equation}
T_{\alpha \beta} N^\alpha N^\beta=\left(\partial\Phi/\partial t+
(1-M/r)^2\partial\Phi/\partial r\right)^2
\end{equation}
and, after substitution of $\Phi$ as in Eq.~(\ref{term})(with M
equal to Q), it is plain to see that our former conclusion
concerning the non-extreme case is recovered.

\section{Disturbing A Static Black Hole in a de Sitter Universe}
\label{d-S}
As a last example of processes in which the horizon area of a static
black hole is left unchanged by an external scalar disturbance, we now
consider the case in which scalar charges are distributed in
a de Sitter universe with a Schwarzschild  black hole
embedded inside it. This spacetime can be described by the metric
\begin{equation}
ds^2 = - (1-2M/r-{\Lambda\over 3}r^2) dt^2 + (1-2M/r-{\Lambda\over
3}r^2)^{-1} dr^2 + r^2 (d\theta^2 +\sin\theta^2 d\varphi^2)
\label{SEdS}
\end{equation}
where $M$ is the mass of the black hole, and $\Lambda$ is the
cosmological constant. This spacetime is characterized by two
horizons which correspond to the two positive roots of the cubic
equation, $g_{tt}(r)=0$. If $9M^2\Lambda<1$ there are three real
roots. One can be attributed to the black hole horizon,
$r_{BH}=r_{BH}(M,\Lambda)$, while the other one is the cosmological
horizon, $r_C=r_C(M,\Lambda)$. The third root $r_0$ is always
negative and thus unphysical. If $9M^2\Lambda=1$, the two horizons
coalesce and are located at  $r=(3M/\Lambda)^{1/3}$. In the case
where $9M^2\Lambda>1$, we have the unphysical situation in which
the mass of the hole exceeds the energy associated with the
cosmological constant.

The shape of the effective potential that a scalar
converging wave experience when it is propagating in this spacetime is
\begin{equation}
V_\ell(r^*)=\left(1-{2M\over r}-
{\Lambda \over 3}r^2\right)\left({2M\over r^3}
-{2\over 3}\Lambda+{\ell(\ell+1)\over r^2}\right).
\end{equation}
In order to test our hypothesis regarding the adiabatic invariance
of the black hole horizon area, we must consider the propagation of
waves in the region between the black hole horizon and the location
of the maximum of the potential barrier. This is due to an assumption
that the squared frequency of the waves is small compared to the
height of the potential barrier. Thus, waves which propagate
outside this region will have no effect on the black hole's
horizon. This problem was dealt with in \cite{BekBrazil3} and a
generalization to the charged case was presented in the previous
sections. Therefore, we shall concern ourselves only with the
effect of scalar waves on the cosmological horizon $r_{C}$, namely
the effect of the waves in the region between the potential barrier
and the cosmological horizon.

For brevity we consider only the time dependent problem. As usual,
we can solve the scalar field equation by a separation of variables.
Thus, after expanding the time dependence in Fourier modes and the
angular dependence in spherical harmonics, the radial part of the
wave equation can be written in term of the difference of the roots
of the cubic equation $g_{tt}(r)=0$; $\gamma\equiv r_{C}-r_{0}$,
$\delta\equiv r_{C}-r_{BH}$, and the radial location of the
cosmological horizon compared to the radial distance to the
cosmological horizon, namely $\epsilon\equiv r_C-r$,
\begin{equation}
{d\over d \epsilon}
\left((r_C-\epsilon)\Delta{d R\over d\epsilon}\right)
+\left(\ell(\ell+1)+{(r_C-\epsilon)^3\omega^2\over \Delta}\right)R=0
\label{sfeds}
\end{equation}
where
$\Delta\equiv\-\Lambda(\gamma-\epsilon)(\delta-\epsilon)\epsilon/3$.
Note that $\gamma$,$\delta$ and $\epsilon$ are all positive semi
definite by definition.

In light of the previous derivation it is obvious that the problem
has two competing scales; the proximity to the cosmological
horizon, denoted here by $\epsilon$, and the distance between the
two horizons, denoted here by $\delta$. In the beginning we will
assume that the black hole horizon is far from the cosmological
horizon and that the sources are placed in a close distance from
the cosmological horizon, namely $r_{BH} \ll r \leq r_C$. Later we
will consider the possibility that the two horizons are on the
verge of coalescence, $\delta\rightarrow 0$.

Therefore, Eq.~(\ref{sfeds}) can be approximated:
\begin{equation}
\epsilon^2 {d^2 R\over d \epsilon^2} +\epsilon{d R\over
d\epsilon}+\left({3 r_C\omega\over\Lambda\gamma\delta}\right)^2
R\simeq0.
\end{equation}
This is an E\"{u}ler equation with the two independent solutions
$R(\epsilon)=N\exp({\pm \imath\eta\ln\epsilon})$, where $N$ is an
arbitrary constant and $\eta=3 r_C\omega/\Lambda\gamma\delta$. Now,
since the cosmological horizon is situated at $\epsilon=0^+$, the
solution with the plus sign should be discarded while the solution
with the minus sign should be retained in accordance with the fact
that only ingoing waves exist at the one-way membrane $r=r_C$.
Therefore, $\Phi$ consists of terms such as
\begin{equation}
P_\ell(\cos\theta)\cos\psi, \quad\quad \psi\equiv\omega(\eta\ln\epsilon+t)
-m\varphi.
\label{term1}
\end{equation}
It can be easily verified that the above solution does not cause
any divergences in the invariants of the curvature. Hence we can
turn now to the problem of verifying that the increase in the area of
the cosmological horizon is at the same rate as the one found in
the previous examples, namely proportional to $\omega^2$, while
keeping in mind that the perturbation to the metric is proportional
only to $\omega$ (see Sec.~\ref{RN1}). To do so, we just have to
calculate expressions such as  the ones in Eq.~(\ref{TNNrn}) with
$N^\alpha$ appropriate to the new spacetime.
\begin{equation}
T_{\alpha \beta} N^\alpha
N^\beta=\left(\partial\Phi/\partial t+(1-2M/r-
{\Lambda\over3}r^2)\partial\Phi/\partial r\right)^2.
\label{TNNds}
\end{equation}
If one now substitutes a $\Phi$ made up of a single mode like in
Eq.~(\ref{term1}), one concludes that
\begin{equation}
T_{\alpha\beta} l^\alpha l^\beta \propto \omega^2 P_\ell^2\sin^2\psi
+\cdots \quad .
\end{equation}
This result agrees with the ones obtained in the previous
situations.

Consider now a series of cases in which the separation between the
black hole horizon and the cosmological horizon $\delta$ is
gradually diminished. As long as $\delta>0$, the potential barrier
maintains its shape, and a {\it weak} scalar wave impinging on the
black hole horizon can have no effect on the cosmological horizon
and {\it vice versa}. If this is so, then our conclusion still
holds: the black hole and cosmological horizons are left unchanged
by the perturbation. In the limit where $\delta\rightarrow 0$ the
potential barrier, which separates the two horizons, disappears and
the wave can either propagate between the innermost side of the
horizon and the singularity, or in the region beyond the
cosmological horizon. Both cases will be left for a future
discussion. It is important to understand that $\delta$ is not a
dynamical parameter in the sense that for a given spacetime
$\delta$ is determined uniquely. However, it has a significant
influence on the magnitude of the sources that may be used in the
adiabatic process. This concludes the example.

\section{Kerr Black Hole Disturbed by an Axially Symmetric
  Array of Scalar Charges}\label{kerr}
In the previous examples we considered a static black hole
disturbed by quasistatic perturbation. Now we shall study in brief
the possibility that the hole itself is rotating. To simplify the
calculation, we assume that the scalar charges are arranged in a
configuration with cylindrical symmetry. We work in Boyer-Lindquist
coordinates with the metric

\begin{eqnarray}
ds^2 = &-& (1-{2Mr\over\rho^2}) dt^2
-4{Mar\sin\theta^2\over\rho^2}d\phi dt
\nonumber \\
&+&\sin\theta^2 \left(r^2+a^2
+{2Mra^2\over\rho^2}\sin\theta^2\right)d\varphi^2
+{\rho^2\over \Delta} dr^2 + \rho^2 d\theta^2 
\label{Kerr}
\end{eqnarray}
where $\Delta=r^2-2Mr+a^2$ and $\rho^2=r^2+a^2\cos\theta^2$. $M$ is
the mass of the hole and $a$ is its specific angular momentum.

As with the static spacetime case, we begin by writing
down the scalar field equation in Kerr spacetime. The details of the
solution of the scalar wave equation in a Kerr background without any
assumptions on the scalar field can be found in
\cite{Fackerell}. Assuming that the scalar field is time independent
and has axial symmetry, one obtains an equation which is
remarkably similar to the equation derived for the scalar field in the
static spherical spacetime. The arguments we used in the static
spherical  spacetime are still applicable here. Therefore, we just
write the {\it physical} solution
\begin{equation}
\Phi = \Re \sum_{\ell=0}^\infty\, C_{\ell}\,
P_\ell \left({r-M\over\sqrt{M^2-a^2}}\right)\, P_\ell(\cos\theta).
\label{PhiKerr}
\end{equation}
As before, it is possible to use the complex coefficient $C_\ell$
in the matching of the solution to any charge configuration with
 said symmetry.

Now, in order to draw a conclusion regarding the adiabatic invariance
of the horizon area, we look at the term $T_\alpha^\beta
N^\alpha N_\beta$ where $N^\alpha$ is proportional to the tangent of
the null generator of the horizon which in Boyer-Lindquist coordinates
reads \cite{MTW}
\begin{eqnarray}
N^t&=&1,
\nonumber \\
N^r&=&{\Delta\over r^2+a^2},
\nonumber \\
N^\theta&=&0,
\nonumber \\
N^\varphi&=&{a\over r^2+a^2}.
\label{generator}
\end{eqnarray}
Since $N^\alpha$ is null, namely $N^r N_r=-(N^t N_t+N^\phi N_\phi)$, we
can write
\begin{equation}
T_\beta^\alpha N^\beta N_\alpha=(T_t^t-T_r^r)N^t
N_t+(T_\varphi^\varphi-T_r^r)N^\varphi N_\varphi.
\label{Ttt-Trr..}
\end{equation}
However, because of the symmetries, the terms $T_t^t-T_r^r$ and
$T_\phi^\phi-T_r^r$ are both equal to $-\nabla_r\Phi \nabla^r\Phi$.
Thus, $T_\beta^\alpha N^\beta N_\alpha$ can be reduced to
$(N^r\partial\Phi/\partial r)^2=\left(\partial\Phi/\partial
r\,\Delta/(r^2+a^2)\right)^2$. Since $\Phi$ is analytic on the
horizon, defined by $\Delta(r_{\cal H})=0$, the horizon area is
left unchanged by the perturbation.

The story becomes more interesting when one considers the time
dependence of the sources. As usual, we make a separation ansatz
which is given by the eigenfunctions appropriate for an axially
symmetric and stationary background geometry, namely, $\Phi_{\ell
\omega}=R_{\ell \omega}(r)\Theta_\ell(\theta)e^{\pm \imath\omega t}$.
By the usual argument we find the separated homogeneous equations,
with separation constant $\lambda_\ell$ \cite{Fackerell},
\begin{eqnarray}
\Delta{d\over d r}\left(\Delta{d R_\ell\over dr}\right)+
(r^2+a^2)^2\omega ^2 R_\ell&=&(\lambda_\ell+\omega^2
a^2\Delta ) R_\ell
\nonumber \\
{1\over\sin\theta}{d\over d
\theta}\left(\sin\theta{d\Theta_\ell\over d
\theta}\right)+a^2\omega^2\cos\theta^2\Theta_\ell
&=&-\lambda_\ell\Theta_\ell.
\label{Phit_kerr}
\end{eqnarray}
The second equation is the same as the flat-space angular spheroidal
equation with $m=0$. Its eigenvalues are denoted by $\lambda_\ell$
and its eigenfunctions $S_\ell(-\imath a\omega,\cos\theta)$. These
eigenfunctions form a discrete set and go over into Legendre
polynomials in the limit $a\omega\rightarrow 0$. The integer $\ell$
has its standards range, but the eigenvalues $\lambda_\ell$ cannot be
analytically expressed in terms of $\ell$ \cite{MathewsWalker}.

The radial equation can be written as a one-dimensional equation with
an effective potential by defining a new radial function and a
new radial coordinate,
\begin{equation}
H_{\ell\omega}(r^*)=\sqrt{r^2+a^2}R_\ell(\omega,r), \quad \quad dr^*=
{r^2+a^2\over\Delta}dr.
\end{equation}

We then find
\begin{eqnarray}
&& \left(-{d^2\over dr^{*2}}+V_{\ell,\omega}(r^*)\right)
H_{\ell\omega}
=\omega^2H_{\ell\omega}
\nonumber \\
&& V_{\ell,\omega}(r^*)\equiv{\Delta\over
(r^2+a^2)^2}\left((\lambda_\ell+\omega^2a^2)+{3r^2-4Mr+a^2\over
r^2+a^2}-{3r^2\Delta\over(r^2+a^2)^2}\right).
\label{u_kerr}
\end{eqnarray}
This potential depends nontrivially on the energy $\omega^2$ of the
field. Now, from Eq.~(\ref{u_kerr}) it is obvious that in the
proximity of the horizon, defined by $\Delta(r_{\cal H})=0$, the
potential is negligible with respect to the ``energy'' of the field
$\omega^2$. Accordingly, we can solve Eq.~(\ref{u_kerr}) to first
order in $|V|/\omega^2$, and write the solution
\begin{equation}
H_{\ell\omega}=\exp(\pm\omega r^*)+\left(\textrm {higher order
terms}\right).
\label{Hkerr}
\end{equation}
We choose appropriate boundary conditions for the solution of
Eq.~(\ref{u_kerr}), analogously to Matzner's choice
\cite{Matzner} for the Schwarzschild case $(a=0)$. Thus, the minus
sign in the $\omega$-mode expansion of $\Phi$ and in Eq.~(\ref{Hkerr})
is singled out. This is a result of the demand that only ingoing
waves exist at the one-way membrane $r^*\rightarrow -\infty$.

Before undertaking the task of verifying that the perturbation does
leave the horizon area unchanged, we have to verify that the
solution is indeed physical, namely invariants of the curvature are
bounded on the horizon. To do so we look, as before, at terms such as 
$\nabla_\alpha\Phi\nabla^\alpha \Phi$. A direct calculation of the
last term delivers the following result:

\begin{equation}
\nabla_\alpha\Phi\nabla^\alpha \Phi ={1\over \rho^2(r^2+a^2)}
\left(a^2\omega^2\sin\psi^2\sin\theta^2 {S_{\ell 0}}^2+\omega r \sin
2\psi S_{\ell 0} +\cos\psi^2 \left({\partial S_{\ell 0} \over
\partial \theta}\right)^2\right)+\cdots\quad .
\end{equation}
where $\psi=\omega(r^*+t)$. Surely, the last result indicates that the
solution we have found does not cause any singularities. Thus, the
solution can be taken to be physical.

Finally, following the chain of reasoning that led us to the
derivation of Eq.~(\ref{Ttt-Trr..}), we find that
\begin{equation}
T_{\alpha\beta}l^\alpha l^\beta =
( l^t\partial\Phi/\partial t+l^r\partial
\Phi/\partial r )^2 \propto
{\omega^2 S_\ell^2 \sin(\psi)^2\over r^2+a^2}+\cdots
\end{equation}
This finding is in accordance with the results obtained in the
previous sections. Thus we find yet another example of the
adiabatic invariance of the black hole horizon area.

\section{Black Hole Disturbed by Electric Charges}\label{electro}
Here we shall examine the influence of an electromagnetic radiation
on the black hole horizon area. In order to sidestep the problems
of electromagnetic interactions between the black hole and the
charges which produce the electromagnetic waves, we assume that the
black hole is neutral. In the first subsection we formulate the
problem. We make use of the possibility to describe the
Schwarzschild space-time as a conformally flat spacetime, occupied
by a ``medium'' which has a non-trivial permeabilities. In the 
following two last subsections we discuss the effect of the
electromagnetic radiation on the horizon area in the static case
and in the dynamic case.

\subsection{Formulation of the problem}
Following the footsteps of \cite{Mashhoon}, Maxwell's equations can
be cast into a simple form if we write the Schwarzschild metric in
isotropic coordinates. Consider the coordinate transformation
$\rho$ defined by $r \equiv M+\rho+M^2/4\rho$ and let
$f(\rho)\equiv 1-2M/r(\rho)$ and $u(\rho)\equiv r(\rho)/\rho$, then
the metric reads
\begin{equation}
ds^2=-f(\rho)dt^2+u(\rho)^2(d\rho^2+\rho^2 d\theta^2+\rho^2
sin\theta^2 d\varphi^2).
\label{isotropic}
\end{equation}
As far as the electromagnetic phenomena are concerned we can think of
the space-time as Minkowskian, occupied by a ``medium'' which is
characterized by dielectric and magnetic permeability tensors
$\epsilon_{ik}=\mu_{ik}\equiv n(\rho) \delta_{ik}$ where
\begin{equation}
n(\rho)\equiv u(\rho)/\sqrt{f(\rho)}.
\label{index}
\end{equation}
$n(\rho)$ can be viewed as an index of refraction induced by
gravity. Note that $n(\rho)$ is everywhere positive definite, and
diverges as $r\rightarrow 2M$. Maxwell's equations can then be
written as
\begin{eqnarray}
{1\over \imath} \overrightarrow{\nabla}\, \times \,
\overrightarrow{F}^\pm &=& \pm n {\partial\overrightarrow{F}^\pm\over
\partial t},
\nonumber \\
\overrightarrow{\nabla}\cdot \,(n \overrightarrow{F}^\pm )&=&0,
\label{Maxwell}
\end{eqnarray}
where
$\overrightarrow{F}^\pm=\overrightarrow{E}\pm\imath\overrightarrow{H}$.
$\overrightarrow{E}$ and $\overrightarrow{H}$ are the electric and
magnetic fields in the medium respectively. The constitutive
equations can be written as
\begin{eqnarray}
\overrightarrow D=n(\rho)\overrightarrow E,
\nonumber \\
\overrightarrow B=n(\rho)\overrightarrow H.
\label{constitutive}
\end{eqnarray}
Note that due to the static nature of the spacetime these constitutive
equations do not include ``gyroscopic'' terms such as $\overrightarrow G
\times \overrightarrow E$ or $\overrightarrow G \times
\overrightarrow H$, where $\overrightarrow G$ represents the
vorticity of the timelike killing vector. $\overrightarrow\nabla$
is taken in the last equation as the usual gradient operator in
3-dimensions, $\hat\rho\,\partial/\partial\rho+\hat\theta
(1/\rho)\,\partial/\partial\theta+\hat\varphi(1/\rho\sin\theta)
\,\partial/\partial\varphi$. $\hat\rho$, $\hat\theta$ and $\hat\varphi$
are unit vectors in the radial, polar and azimuthal directions
respectively.

Consider now the representation
\begin{equation}
\overrightarrow{F}^\pm(\overrightarrow{\rho},t)=\sum_{k}
\int d\omega \, e^{-\imath \omega t} \sum_{\sigma=0,e,\mu}\,
F_{k}^{\pm(\sigma)}(\rho,\omega)
\overrightarrow{Y}^{(\sigma)}_{k}(\hat{\rho}),
\label{rep}
\end{equation}
$k$ is an alias for $(\ell,m)$. $\ell$ and $m$ have the usual
range. $\overrightarrow{Y}^{(e)}_{k}$ and
$\overrightarrow{Y}^{(\mu)}_{k}$ are the transverse and
$\overrightarrow{Y}^{(0)}_{k}$ the longitudinal vector spherical
harmonics \cite{VSH}. They may be expressed by
\begin{eqnarray}
\overrightarrow Y_{k}^{(0)}&=&\hat \rho \, Y_{k}(\hat \rho),
\nonumber \\
\overrightarrow Y_{k}^{(e)}&=&\overrightarrow\nabla_{\hat \rho} \,
Y_{k}(\hat \rho),
\nonumber \\
\overrightarrow Y_{k}^{(\mu)}&=&\hat \rho \times
\overrightarrow\nabla_{\hat \rho} \, Y_{k}(\hat \rho).
\label{vsh}
\end{eqnarray}
When one substitutes this in Maxwell's equations,
Eq.~(\ref{Maxwell}), one gets
\begin{eqnarray}
-\sqrt{\ell(\ell+1)}F_{k}^{\pm(\mu)}&=&\pm\omega n(\rho)
F_{k}^{\pm(0)},
\nonumber \\
-{d\over d\rho}\left(\rho F_{k}^{\pm(\mu)}\right)&=&\pm \omega
 \rho n(\rho)F_{k}^{\pm(e)},
\nonumber \\
 {d\over d\rho}\left(\rho F_{k}^{\pm(e)}\right)-\sqrt{\ell
 (\ell+1)} F_{k}^{\pm(0)}&=&\pm\omega\rho n(\rho) F_{k}^{\pm(\mu)},
\nonumber \\
{d F_{k}^{\pm(0)}\over d\rho}+
{d\over d\rho}\left(\ln (\rho^2 n)\right) F_{k}^{\pm(0)}+
\ell(\ell+1) F_{k}^{\pm(e)}&=&0.
\label{maxwell1}
\end{eqnarray}

An indication of the relevance of the solution to the physical
problem at hand can be obtained by examining the invariants that
may be constructed from the energy-momentum tensor of the
perturbation
\begin{equation}
T_\mu^\nu={1\over
  4\pi\sqrt{-g}}\left(F_{\mu\alpha}{\cal
    H}^{\beta\alpha}\delta_\beta^\nu-{1\over 4}\delta_\mu^\nu
  F_{\alpha\beta}{\cal H}^{\alpha\beta}\right),
\label{EMTEM}
\end{equation}
where we define the usual field strength,
$F_{\alpha\beta}\rightarrow (\overrightarrow E,\overrightarrow B)$
and the pseudo-tensor ${\cal H}^{\alpha\beta}\equiv \sqrt{-g}
F^{\alpha\beta} \rightarrow (-\overrightarrow D,\overrightarrow
H)$. Note that $\sqrt{-g}=u^4/n$. Every invariant of this kind must
be bounded for a physically acceptable solution. The analysis can
be simplified much  using the following arguments. $T^\nu_\mu$ may
be seen as a $4\times 4$ real matrix \cite{BekMayo}. Thus, any
invariant constructed from $T^\nu_\mu$ is actually the trace of
some power of ${\bf T}$. For example, $T^\alpha_\alpha={\rm Tr}\,{\bf
T}\equiv 0$, $T^\alpha_\beta T^\beta_\alpha={\rm Tr}\,{\bf T}^2$,
$T^\alpha_\beta T^\beta_\gamma T^\gamma_\alpha={\rm Tr}\,{\bf T}^3$ 
etc. Now, in this matrix notation $T^\nu_\mu$ can be written as
\begin{equation}
{\bf T}={1\over
  4\pi\sqrt{-g}}\left({\bf M}-{1\over 4} {\rm Tr} {\bf M}\,
  {\textrm {\rm\bf\large 1}}\right)
\end{equation}
where $M^\nu_\mu\equiv F_\mu^\alpha {\cal H}^\nu_\alpha$ or in
matrix notation ${\bf M}\equiv {\bf F}\cdot {\bf\cal H}$ and
{\rm\bf\large 1} is a $4\times 4$ identity matrix. The trace of any
power of ${\bf T}$ is simply the trace of some polynomial function
of ${\bf M}$, where the polynomials coefficients may be powers of
${\rm Tr}\,{\bf M}$. Now, since the trace operator is linear, it is
sufficient to look at expressions like ${\rm Tr}\,{\bf M},\, {\rm Tr}\,{\bf
M}^2,\,{\rm Tr}\,{\bf M}^3$ {\it etc}. These in turn are simply the sum
of the eigenvalues of ${\bf M}$ raised to the appropriate power.
Hence, a prerequisite for the boundness of the invariants of the
curvature is that the eigenvalues of ${\bf M}$ vanishe at least as
fast as $1/n$ as we approach the horizon. It turns out that ${\bf
M}$ has only two distinct eigenvalues which have each a two-fold
degeneracy. These are
\begin{equation}
\lambda_1={1\over 4}\left(\|{\overrightarrow F}^+\|+\|{\overrightarrow
  F}^-\|\right)^2,  \quad\quad\quad
\lambda_2={1\over 4}\left(\|{\overrightarrow F}^+\|-\|{\overrightarrow
  F}^-\|\right)^2
\end{equation}
where $\|{\overrightarrow F}^\pm\|\equiv\left({\overrightarrow
F}^\pm\cdot{\overrightarrow F}^\pm\right)^{1/2}$. Now if one uses
the definition of the vector spherical harmonic Eq.~(\ref{vsh}) it
is plain to see that $\overrightarrow Y_{k_1}^{(e)} \cdot
\overrightarrow Y_{k_2}^{(e)}=\overrightarrow Y_{k1}^{(\mu)} \cdot
\overrightarrow Y_{k_2}^{(\mu)}$ for every $k$, and  since the
vector spherical harmonics form an orthogonal base,
$\|{\overrightarrow F}^\pm\|$ can be reduced to
\begin{eqnarray}
\|{\overrightarrow F}^\pm\|^2 &=&\Re
\sum_{k_1,k_2} \, \int d\omega_1 d\omega_2
e^{-\imath(\omega_1+\omega_2)t}
\nonumber \\
&\,& \left(Y_{k_1}Y_{k_2}F^{\pm(0)}_{k_1} F^{\pm(0)}_{k_2}+
\overrightarrow \nabla Y_{k_1} \cdot \overrightarrow \nabla Y_{k_2}
\sum_{\sigma=e,\mu}\,F^{\pm(\sigma)}_{k_1} F^{\pm(\sigma)}_{k_2}
\right).
\label{FH}
\end{eqnarray}
In the following sections we will use this expression to verify that
our solution is physically meaningful.

After the solution is shown to be physical, we will test whether
the horizon area is left unchanged under the disturbance. This will
be confirmed if the term $4\pi T_{\alpha\beta} l^\alpha l^\beta$
vanishes on the horizon. In Sec.~\ref{RN1} $l^\alpha$, the null
generator of the horizon was shown to be proportional to the
vector $N^\alpha$. Now, under the coordinates transformation to the
isotropic coordinates system, the vector  $N^\alpha$ transforms to
\begin{equation}
N^\alpha\equiv(1,1-2M/r,0,0)\rightarrow(1,1/n,0,0).
\end{equation}
Thus, a straightforward calculation leads to the following result:
\begin{equation}
4\pi T_{\alpha\beta} l^\alpha l^\beta\propto {1\over u^2} \Re
\left({F_\theta}^+ +\imath {F_\phi}^+\right)\left({F_\theta}^- -\imath
{F_\phi}^-\right)
\label{TNN}
\end{equation}
and after substitution of $\overrightarrow F^\pm$ from Eq.~(\ref{rep}),
one obtains
\begin{equation}
{F_\theta}^\pm \pm\imath {F_\phi}^\pm=
\sum_{k}\int d\omega {e^{-\imath\omega t}\over \rho}
\left(F_{k}^{\pm(e)}\pm\imath F_{k}^{\pm(\mu)}\right){\cal D}^\pm Y_{k}
\label{F+F}
\end{equation}
where ${\cal D}^\pm$ is the differential operator
$\partial/\partial\theta\pm
(\imath/\sin\theta)\partial/\partial\varphi$. All the above is
general. In the following subsection we shall discuss the static
problem. Finally, in the last subsection we shall retain the time
dependence of the fields and solve the dynamical problem.

\subsection{The Static Problem}
When the electromagnetic fields produced by the charges are static,
Maxwell's equations [Eq.~(\ref{maxwell1})] can be reduced to
\begin{eqnarray}
\sqrt{\ell(\ell+1)}F_{k}^{\pm(\mu)}&=&0,
\nonumber \\
{d\over d\rho}\left(\rho F_{k}^{\pm(\mu)}\right)&=&0,
\nonumber \\
{d\over d\rho}\left(\rho F_{k}^{\pm(e)}\right)-\sqrt{\ell
  (\ell+1)} F_{k}^{\pm(0)}&=&0,
\nonumber \\
{d F_{\ell m}^{\pm(0)}\over d\rho}+{d\over
    d\rho}\left(\ln (\rho^2 n)\right)F_{k}^{\pm(0)}+
    \ell(\ell+1) F_{k}^{\pm(e)}&=&0.
\label{maxwell_static}
\end{eqnarray}
The first two equations simply state that $F_{k}^{\pm(\mu)}=0$
for every $k$. The last two equations can be combined to give
\begin{eqnarray}
 {d^2\over d\rho^2}\left(\rho F_{k}^{\pm(e)}\right)&+&{d\over
    d\rho}\left(\ln (\rho^2 n)\right){d\over d\rho}
    \left(\rho F_{k}^{\pm(e)}\right)+
\left(\ell(\ell+1)\right)^{3/2}F_{k}^{\pm(e)}=0
\nonumber \\
 F_{k}^{\pm(0)}&=&{1\over\sqrt{\ell
  (\ell+1)}}{d\over d\rho}\left(\rho F_{k}^{\pm(e)}\right).
\label{F0Fe}
\end{eqnarray}
$F_{k}^{\pm(0)}$ is given by the second equation of Eq.~(\ref{F0Fe}) after
the first equation is solved for $F_{k}^{\pm(e)}$.

After the substitution of $n(\rho)$ and the introduction of a new
function $X_{k}^{\pm(e)}=\rho F_{k}^{\pm(e)}$, we can write the
equation for $F_{k}^{\pm(e)}$ in the following form
\begin{equation}
{d^2 X_{k}^{\pm(e)}\over d\rho^2}+2{\rho-M\over (\rho+{M\over2})
(\rho-{M\over2})}{d X_{k}^{\pm(e)}\over
    d\rho}+\left(\ell(\ell+1)\right)^{3/2}{X_{k}^{\pm(e)}\over \rho}=0.
\label{eq}
\end{equation}
Subsequently we will be interested in the behavior of the
electromagnetic fields in the proximity of the horizon. Therefore,
we look for an approximate solution to Eq.~(\ref{eq}) near
$\rho=M/2$. We introduce a new variable $x=\rho-M/2$ and a new
function $X_{k}^{\pm(e)}(x)=x z_{k}^{\pm(e)}(x)$. Then
Eq.~(\ref{eq}) assumes the following form
\begin{equation}
x^2 {d^2 z_{k}^{\pm(e)}\over dx^2}+x {d z_{k}^{\pm(e)}\over dx}
+\left(\alpha x^2-1\right) z_{k}^{\pm(e)}\simeq0
\end{equation}
with $\alpha=2 \left(\ell(\ell+1)\right)^{3/2}/M$. This equation is the
familiar Bessel equation of order $1$ with the two independent
solutions $J_1 (\sqrt{\alpha} x)$ and $Y_1(\sqrt{\alpha} x)$.
Convergence of the curvature invariant, Eq~(\ref{FH}), forbids us
from considering the $Y_1(\sqrt{\alpha} x)$ solution as a physical
one since it diverges at $x=0$, namely at the horizon. As in the
former cases, this divergence is unacceptable since it will induce
divergences in the invariants of the curvature, thus disqualifying
the perturbation method.

Using Eq.~(\ref{F0Fe}), it is now possible to show that
\begin{eqnarray}
F_{k}^{\pm(\mu)}&\equiv&0,
\nonumber \\
F_{k}^{\pm(e)}&\simeq&{x\over \rho}C_{k}^\pm J_1(\sqrt{\alpha}x),
\nonumber \\
F_{k}^{\pm(0)}&\simeq&{C_{k}^\pm\over
  \sqrt{\ell(\ell+1)}}\left(J_1(\sqrt{\alpha}x)-{\sqrt{\alpha}x\over 2}
  \left(J_0(\sqrt{\alpha}x)-J_2(\sqrt{\alpha}x)\right)\right).\\
\nonumber
\end{eqnarray}
Hence, according to Eq.~(\ref{TNN}) and Eq.~(\ref{F+F}) (with
$\omega=0$) the behavior of $4\pi T_{\alpha\beta} l^\alpha l^\beta$
is governed solely by the behavior of $F_{k}^{\pm(e)}$ near the
horizon. Now, for $x\rightarrow 0$, $J_1(\sqrt{\alpha} x)$ is of
 $O(x)$. Therefore, $F_{k}^{\pm(e)}$ is of $O(x^2)$ and
$F_{k}^{\pm(0)}$ is of $O(x)$. This implies that $4\pi
T_{\alpha\beta} l^\alpha l^\beta$ is of order $O(x^4)$. Thus the
horizon area does not change under the action of the static
electromagnetic fields.

Since the adiabatic invariance of an object is primarily meaningful
in the context of dynamic processes, we turn now to deal with the
dynamic problem. Basically, we will consider the process in which
the charges are brought adiabatically from infinity through the
potential barrier of the black hole into the immediate surroundings
of the horizon. This will be dealt with in the next subsection.

\subsection{Moving the Charges Around}
First we introduce a new coordinate $r^*$ defined by
$dr^*/d\rho\equiv n(\rho)$. Then $r^*(\rho)$ can be identified with
Wheeler's ``tortoise'' coordinate \cite{MTW};
$r^*=r+2M\ln{(r/2M-1)}$. Moreover, as usual, we define new radial
functions $X_{k}^{\pm (\sigma)}(r^*,\omega)
=\rho F_{k}^{\pm(\sigma)}$. In terms of these new radial functions
the radial parts of Maxwell's equations reads
\begin{eqnarray}
-{d^2X_{k}^{\pm (\mu)}\over d {r^*}^2} &+& \left(1-{2 M\over
      r}\right) {\ell(\ell+1)\over r^2}X_{k}^{\pm (\mu)}=\omega^2
  X_{k}^{\pm (\mu)},
\nonumber \\
X_{k}^{\pm (e)}&=&\mp(1/\omega) {d X_{k}^{\pm (\mu)}\over dr^*},
\nonumber \\
X_{k}^{\pm (0)}&=&\mp(1/\omega)\sqrt{\ell(\ell+1)} {X_{k}^{\pm
    (\mu)}\over \rho n}.
\label{maxwell2}
\end{eqnarray}

We are interested in a solution in the region where $\omega^2$
is much larger than the potential barrier, namely in the region where
\begin{equation}
\omega^2\gg \left(1-{2 M\over r}\right) {\ell(\ell+1)\over r^2}.
\label{gg}
\end{equation}
An approximate solution is
\begin{eqnarray}
X_{k}^{\pm (\mu)}&\simeq& C^\pm_{k}(\omega) e^{-\imath\omega r^*},
\nonumber \\
X_{k}^{\pm (e)}&\simeq& \pm \imath C^\pm_{k}(\omega)
e^{-\imath\omega r^*},
\nonumber \\
X_{k}^{\pm (0)}&\simeq& \mp \sqrt{\ell(\ell+1)}
  {C^\pm_{k}(\omega) e^{-\imath\omega r^*}\over \omega \,\rho\, n}.
\label{approxi}
\end{eqnarray}
A useful by-product of the fact that $1/n$ vanishes on the horizon, is
that the longitudinal component of the electromagnetic fields vanishes
as $\sqrt{1-2M/r}$ when $r\rightarrow 2M$.

After substitution of $X^{\pm(\sigma)}$ from Eq.~(\ref{approxi}) in
Eq.~(\ref{FH}), it is evident that the summation over $\sigma$
vanishes identically, regardless of the magnitude of the frequency
dependent coefficients, $C^\pm_{k}(\omega)$. And as mentioned, the
longitudinal term $F^{\pm(0)}_{k_1}F^{\pm(0)}_{k_2}$  is
proportional to $1/n^2$. Hence, the multiplication of this term by
$n$ does not cause any trouble. In fact as we approach the horizon
the contribution from this term to the perturbation of the metric
becomes negligible. This shows that every invariant of the type we
considered vanishes on the horizon in spite of the divergence of
$n$ there. Of course, globally these invariants may still have an
effect on the geometry. In particular the geometry of the horizon
can get disturbed.

Now, since the perturbation is supposed to be weak, we are limited to
small frequencies, namely long wavelengths compared to the size of
the hole, $\lambda>>2M$. Additionally the coefficients
$C^\pm_{k}(\omega)$  must be bounded everywhere and are assumed to be
 smooth functions of $\omega$. As before, consider a solution
composed of a single mode. Then, substitution of this solution,
Eq.~(\ref{approxi}), in Eq.~(\ref{FH}) gives
\begin{equation}
\|{\overrightarrow F}^\pm\|^2 \simeq \ell(\ell+1)\Re\left({
   C_k^{\pm}(\omega)\over \omega}\right)^2
\left(P_\ell{\cos\psi\over \rho^2 n } \right)^2=
O\left(\left[{C_k(\omega)\over\omega}\right]^2\right)
\end{equation}
where $\psi=\omega(r^*+t)-m\varphi$. It is evident from above
that at a small radial distance from the horizon the
electromagnetic fields induce an oscillatory disturbance of
$O(\left[C_k(\omega)/\omega\right]^2)$. The disturbance vanishes on
the horizon. The limit $\omega\rightarrow 0$ was dealt with in the
previous subsection and was found to exist. Now, how does the
horizon area change when the weak perturbation is turned on?
Consider again a solution composed of a single mode. Then, the
substitution of this solution in Eq.~(\ref{F+F}) and
Eq.~(\ref{TNN}) gives
\begin{equation}
4\pi T_{\alpha\beta} l^\alpha l^\beta\propto
\left({2 \cos\psi\over
  \rho^2 u}\right)^2\left(\left(d P_\ell\over d\theta\right)^2-
  \left(m P_\ell\over \sin\theta\right)^2\right)
\Re\left(C_{k}^{+}(\omega)C_{k}^{-}(\omega)\right)=O(C_k(\omega)^2).
\end{equation}
Obviously, this result states that the perturbation to the horizon
area is a quantity of higher order in $\omega$ than the
perturbation to the geometry as quantified by the invariants
calculated previously. Thus horizon area is again seen to be an
adiabatic invariant.

\medskip

{\bf ACKNOWLEDGMENTS} The author would like to thank Professor
J. D. Bekenstein for proposing this problem and for his suggestions
and advice, and to S. Hod for helpful discussions. This work is
supported by a grant from the Israel Science Foundation established
by the Israel Academy of Sciences.

\end{document}